\documentclass[twocolumn,amsmath,amssymb,aps,prb]{revtex4-1}  % 2-column
\usepackage{graphicx}           % for arxiv
\usepackage{dcolumn}
\usepackage{color}
\usepackage{bm}      
\usepackage{xspace}

\newcommand{\be}{\begin{equation}}
\newcommand{\ee}{\end{equation}}
\newcommand{\bea}{\begin{eqnarray}}
\newcommand{\eea}{\end{eqnarray}}
\newcommand{\beaa}{\begin{eqnarray*}}
\newcommand{\eeaa}{\end{eqnarray*}}

\begin{document}

\title{
    Spin-Spin Correlations in the Kitaev Model at Finite Temperatures:
    Approximate and Exact Results via Green's Function Equation of Motion
}

\author{Hibiki Takegami}
\email{takegami.hibiki.64h@st.kyoto-u.ac.jp}
 \affiliation{
   Course of Studies on Materials Science, 
  Graduate School of Human and Environmental Studies, 
  Kyoto University, Kyoto 606-8501, Japan
}
\author{Takao Morinari}
 \email{morinari.takao.5s@kyoto-u.ac.jp}
 \affiliation{
   Course of Studies on Materials Science, 
  Graduate School of Human and Environmental Studies, 
  Kyoto University, Kyoto 606-8501, Japan
}

\date{\today}

\begin{abstract}
  The Kitaev model, defined on a honeycomb lattice, features an exactly solvable ground state
  with fractionalized Majorana fermion excitations,
  which can potentially form non-Abelian anyons crucial for fault-tolerant topological quantum computing.
  Although Majorana fermions are essential for obtaining the exact ground state,
  their physical interpretation in terms of spin operators remains unclear.
  In this study, we employ a Green's function approach that maintains SU(2) symmetry
  to address this issue and explore the model's finite temperature properties.
  Our results demonstrate that the computed temperature dependence of the correlation functions
  closely approximates the exact values at zero temperature, confirming the accuracy of our method.
    We also present several exact results concerning the spin Green's function and
    spin-spin correlation functions that are specific to the Kitaev model.
\end{abstract}

\maketitle

Quantum spin liquids have emerged as a central topic in condensed matter physics,
capturing significant attention due to their complex and intriguing properties \cite{Savary2017}.
Among the various theoretical models, the Kitaev model is particularly notable
for its exactly solvable ground state, which features fractionalized Majorana fermions \cite{Kitaev2006}.
This model is not only analytically tractable but also holds potential for real-world applications
in materials with significant spin-orbit coupling \cite{Jackeli2009, Takagi2019}.
Majorana fermions play a crucial role in the development of topological quantum computers,
offering a platform for fault-tolerant quantum computation \cite{Kitaev2003, Nayak2008}.

While investigating the Kitaev model in terms of Majorana fermions provides
significant mathematical elegance, a deeper understanding of the physical nature of Majorana fermions
may be achieved through the study of spin operators.
However, the Kitaev model presents substantial challenges due to its highly frustrated
and strongly correlated nature.

In this paper, we employ the spin Green's function approach for the spin operators
while preserving SU(2) symmetry and solve its equation of motion.
This method, which maintains SU(2) symmetry \cite{Kondo1972},
is particularly suitable because the system does not exhibit magnetic long-range order
even at zero temperature.
Another critical observation is that the spin-spin correlation function in the ground state is
finite only between nearest neighbor sites \cite{Baskaran2007},
suggesting that the spin-spin correlation length remains short-ranged even at finite temperatures.
Since the equation of motion for the Green's function \cite{Zubarev1960} relies on
finite-range correlation functions, it is well-suited for exploring
the finite temperature properties of the Kitaev model.
Additionally, this approach aligns with the high-temperature expansion \cite{Kondo1972},
enhancing the accuracy of results at higher temperatures.
Physically, our spin Green's function describes the propagation of a pair of $\mathbb{Z}_2$ fluxes.
Our results demonstrate that the computed temperature dependence of the correlation functions
yields a value at zero temperature that is quite close to the exact value.
  Furthermore, we present several exact results regarding spin-spin correlation functions
  at finite temperatures.

\begin{figure}[htbp]
  \includegraphics[width=0.7 \linewidth, angle=0]{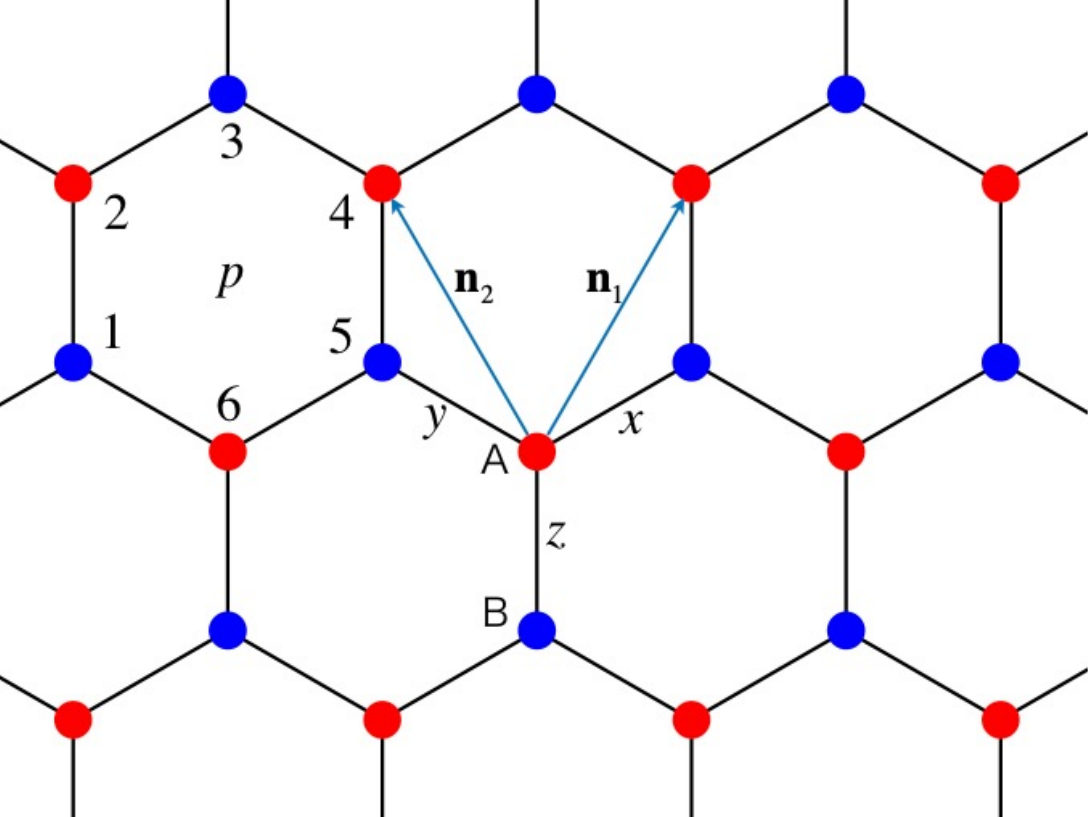}
  \caption{
    \label{fig:honeycomb_lattice}
    (Color online)
    Kitaev model on the honeycomb lattice, showing two interpenetrating sublattices, A and B.
    Sublattice A is represented by red circles, while sublattice B is represented by blue circles.
    The direction-dependent interactions in Eq.~(\ref{eq:H}) are labeled as $x$, $y$, and $z$.
    The lattice vectors are denoted as
    ${\bf n}_1 = (\sqrt{3}/2, 3/2)a_0$ and ${\bf n}_2 = (-\sqrt{3}/2, 3/2)a_0$,
    where $a_0$ is the nearest neighbor distance.
    The plaquette operator $W_p$ for plaquette $p$ is defined by Eq.~(\ref{eq:Wp}).
  }
\end{figure}

The Hamiltonian of the isotropic Kitaev model on the honeycomb lattice is given by \cite{Kitaev2006}:
\be
   {\cal H} =
     -
   K \sum_{\gamma = x,y,z} \sum_{{\langle i,j \rangle}_\gamma} S_i^\gamma S_j^\gamma,
\label{eq:H}
\ee
where $S_i^\gamma$ is the $\gamma$-component of the spin-1/2 operator at site $i$,
  and we consider the ferromagnetic Kitaev coupling $K>0$.
The summation of ${\langle i,j \rangle}_\gamma$ represents
the sum over the nearest neighbor sites connected by a $\gamma$ bond,
as shown in Fig.~\ref{fig:honeycomb_lattice}.
In the honeycomb lattice, each unit cell comprises two sites corresponding to the A and B sublattices,
as illustrated in Fig.~\ref{fig:honeycomb_lattice}.
Additionally, Fig.~\ref{fig:honeycomb_lattice} shows a plaquette $p$ consisting of six sites.
For this plaquette, we define the following plaquette operator \cite{Kitaev2006}:
\be
W_p = \sigma_1^x \sigma_2^y \sigma_3^z \sigma_4^x \sigma_5^y \sigma_6^z,
\label{eq:Wp}
\ee
where $\sigma_i^{\gamma}$ represents the $\gamma$-component of the Pauli matrix at site $i$.
  $\sigma_i^{\gamma}$ can be expressed as $\sigma _i^\gamma = ib_i^\gamma c_i$
  in terms of two types of Majorana fermions \cite{Kitaev2006},
  where this equality holds by restricting the extended Hilbert space of Majorana fermions
  to the physical Hilbert space of the original spin operators.
  The product of two Majorana fermions, $b_i^\gamma$ and $b_j^\gamma$,
  residing at nearest neighbor sites
  forms a $\mathbb{Z}_2$ gauge field.
  In this formulation, $W_p$ represents the $\mathbb{Z}_2$ gauge flux.
  Meanwhile, the Majorana fermions $c_i$ are itinerant under these $\mathbb{Z}_2$ gauge fields.

%------------------------------------------------------------------------  
%------------------------------------------------------------------------

An intriguing aspect of the Kitaev model is that it has been rigorously demonstrated
that its ground state is a spin liquid state without long-range magnetic order.
Because of the absence of the magnetic long-range order,
we may apply SU(2) invariant formalsim of the Green's function method\cite{Kondo1972}.
We define the Matsubara Green's function:
\be
G_{n{\alpha _1},m{\alpha _2}}^\gamma \left( \tau  \right)
=  - \left\langle {{T_\tau }
  S_{n{\alpha _1}}^\gamma \left( \tau  \right)
  S_{m{\alpha _2}}^\gamma \left( 0 \right)} \right\rangle
\equiv
{\left\langle {{S_{n{\alpha _1}}^\gamma }}
 \mathrel{\left | {\vphantom {{S_{n{\alpha _1}}^\gamma } {S_{m{\alpha _2}}^\gamma }}}
 \right. \kern-\nulldelimiterspace}
 {{S_{m{\alpha _2}}^\gamma }} \right\rangle _\tau }.
\label{eq:G_spin:real_space}
\ee
Here $\tau$ represents the imaginary time
and $T_{\tau}$ represents the imaginary time ordering operator.
The notation $S_{n{\alpha}}^\gamma$ represents
the $\gamma$-th component of the spin operator
for the $\alpha$-th sublattice within the $n$-th unit cell.

\begin{figure}[htbp]
  \includegraphics[width=0.8 \linewidth, angle=0]{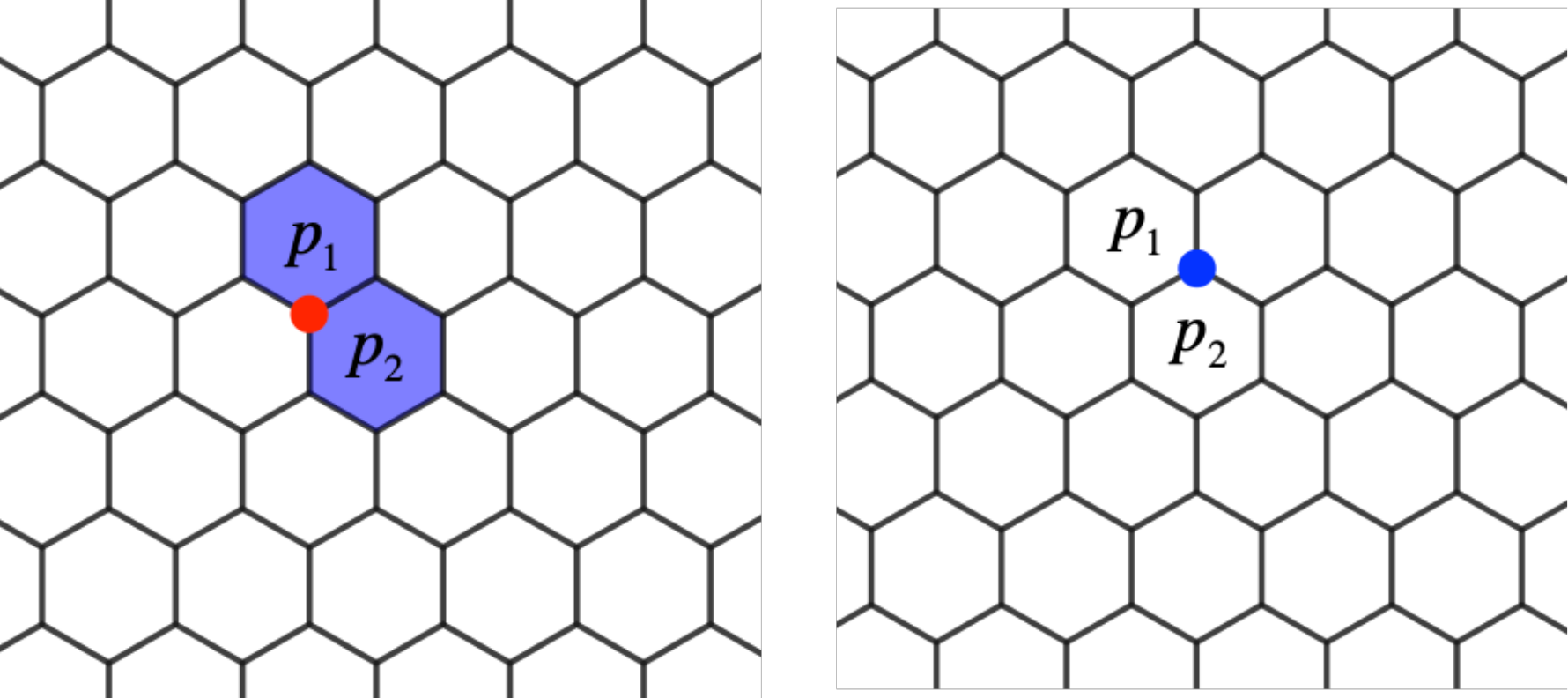}
  \caption{
    \label{fig:GreenFuncZ2Flux}
    (Color online)    
    The physical meaning of the Green's function (\ref{eq:G_spin:real_space})
    for the case of $\gamma = x$, $\alpha_1 = B$, and $\alpha_2 = A$,
    with the $n$-th unit cell shifted from the $m$-th unit cell by ${\bf n}_1$.
    In the left panel, the red circle denotes the A sublattice in the $m$-th unit cell.
    At imaginary time $0$, the $\mathbb{Z}_2$ flux values in plaquettes $p_1$ and $p_2$
    are flipped by the spin operator $S_{m{\rm A}}^x$, as indicated by the filled hexagons.
    In the right panel, the flipped $\mathbb{Z}_2$ flux values
    are flipped again at imaginary time $\tau$ by the spin operator $S_{n{\rm B}}^x$.
  }
\end{figure}

Before delving into the analysis of the Green's function,
we first discuss its physical meaning.
Specifically, we consider the Green's function (\ref{eq:G_spin:real_space})
with $\gamma = x$, $\alpha_1 = B$, and $\alpha_2 = A$.
This Green's function describes a process where the $\mathbb{Z}_2$ flux values
at adjacent hexagon plaquettes are flipped at imaginary time $0$
and then flipped again at imaginary time $\tau$.
This process is schematically illustrated in Fig.~\ref{fig:GreenFuncZ2Flux}.
We observe that any flipped $\mathbb{Z}_2$ flux values must be flipped again;
otherwise, the thermal average will vanish.
%-------------------------------------------------------------------
%-------------------------------------------------------------------

Denoting the inverse temperature by $\beta = 1/T$, where $T$ is the temperature,
and setting the Boltzmann constant $k_{\rm B}$ to 1,
the Fourier representation of Eq.~(\ref{eq:G_spin:real_space})
in the Matsubara frequency, $\omega_n = 2\pi n/\beta$ with $n$ being an integer,
is given by
\bea
G_{n{\alpha _1},m{\alpha _2}}^\gamma \left( {i{\omega _n}} \right)
&=& \int_0^\beta  {d\tau } {e^{i{\omega _n}\tau }}G_{n{\alpha _1},m{\alpha _2}}^\gamma \left( \tau  \right)
\nonumber \\
& \equiv & 
{\left\langle {{S_{n{\alpha _1}}^\gamma }}
  \mathrel{\left |
    {\vphantom {{S_{n{\alpha _1}}^\gamma } {S_{m{\alpha _2}}^\gamma }}}
 \right. \kern-\nulldelimiterspace}
 {{S_{m{\alpha _2}}^\gamma }} \right\rangle _{i{\omega _n}}}
\eea
To find the expression for the Green's function,
we first take the derivative of Eq.~(\ref{eq:G_spin:real_space})
with respect to $\tau$ and then perform the Fourier transform.
This process yields 
   \bea
   i{\omega _n}{\left\langle {{S_{n{\alpha _1}}^\gamma }}
 \mathrel{\left | {\vphantom {{S_{n{\alpha _1}}^\gamma } {S_{m{\alpha _2}}^\gamma }}}
   \right. \kern-\nulldelimiterspace}
     {{S_{m{\alpha _2}}^\gamma }} \right\rangle _{i{\omega _n}}}
   &=& {\left\langle {{\left[ {{\cal H},S_{n{\alpha _1}}^\gamma } \right]}}
 \mathrel{\left | {\vphantom {{\left[ {{\cal H},S_{n{\alpha _1}}^\gamma } \right]} {S_{m{\alpha _2}}^\gamma }}}
   \right. \kern-\nulldelimiterspace}
         {{S_{m{\alpha _2}}^\gamma }} \right\rangle _{i{\omega _n}}}
   \nonumber \\ & & 
   + \left\langle {\left[ {S_{n{\alpha _1}}^\gamma ,S_{m{\alpha _2}}^\gamma } \right]} \right\rangle.
   \eea
   Now we have a new Green's function, represented by the first term on the right-hand side.
   The equation of motion for this Green's function is given by   
   \bea
   i{\omega _n}{\left\langle {{\left[ {{\cal H},S_{n{\alpha _1}}^\gamma } \right]}}
 \mathrel{\left | {\vphantom {{\left[ {{\cal H},S_{n{\alpha _1}}^\gamma } \right]} {S_{m{\alpha _2}}^\gamma }}}
 \right. \kern-\nulldelimiterspace}
         {{S_{m{\alpha _2}}^\gamma }} \right\rangle _{i{\omega _n}}}
   &=& {\left\langle {{\left[ {{\cal H},\left[ {{\cal H},S_{n{\alpha _1}}^\gamma } \right]} \right]}}
 \mathrel{\left | {\vphantom {{\left[ {{\cal H},\left[ {{\cal H},S_{n{\alpha _1}}^\gamma } \right]} \right]} {S_{m{\alpha _2}}^\gamma }}}
 \right. \kern-\nulldelimiterspace}
         {{S_{m{\alpha _2}}^\gamma }} \right\rangle _{i{\omega _n}}}
   \nonumber \\
   & & + \left\langle {\left[ {\left[ {{\cal H},S_{n{\alpha _1}}^\gamma } \right],S_{m{\alpha _2}}^\gamma } \right]} \right\rangle.
   \eea
   We first compute $\left[ {{\cal H},S_{n{\alpha 1}}^\gamma } \right]$,
   and then compute $\left[ {{\cal H},\left[ {{\cal H},S_{n{\alpha _1}}^\gamma } \right]} \right]$.
   From the latter calculation, terms like 
   ${\left\langle {{S_{nA}^\gamma S_{{n_1}B}^xS_{{n_2}A}^\gamma }}
 \mathrel{\left | {\vphantom {{S_{nA}^\gamma S_{{n_1}B}^xS_{{n_2}A}^\gamma } {S_{m\alpha }^x}}}
 \right. \kern-\nulldelimiterspace}
         {{S_{m\alpha }^x}} \right\rangle _{i{\omega _n}}}$,
   ${\left\langle {{S_n^\gamma S_{{n_1}B}^\gamma S_{{n_2}A}^x}}
 \mathrel{\left | {\vphantom {{S_n^\gamma S_{{n_1}B}^\gamma S_{{n_2}A}^x} {S_{m\alpha }^x}}}
 \right. \kern-\nulldelimiterspace}
         {{S_{m\alpha }^x}} \right\rangle _{i{\omega _n}}}$,
   etc., appear.
   Instead of considering the equation of motion for such terms,
     we approximate them as the product of a correlation function
     and a Green's function with a reduced number of spins,
     using the so-called Tyablikov decoupling \cite{Tyablikov1962},
     as follows:
   \bea
   {\left\langle {{S_{nA}^\gamma S_{{n_1}B}^xS_{{n_2}A}^\gamma }}
 \mathrel{\left | {\vphantom {{S_{nA}^\gamma S_{{n_1}B}^xS_{{n_2}A}^\gamma } {S_{m\alpha }^x}}}
 \right. \kern-\nulldelimiterspace}
         {{S_{m\alpha }^x}} \right\rangle _{i{\omega _n}}}
   &\simeq & \alpha c_{AA}^\gamma \left( {{{\bf{R}}_{{n_2}}} - {{\bf{R}}_n}} \right)
   \nonumber \\ & & \times
   {\left\langle {{S_{{n_1}B}^x}}
     \mathrel{\left | {\vphantom {{S_{{n_1}B}^x} {S_{m\alpha }^x}}}
       \right. \kern-\nulldelimiterspace}
             {{S_{m\alpha }^x}} \right\rangle _{i{\omega _n}}},
   \label{eq:decoupling1}
   \\
     {\left\langle {{S_{nA}^\gamma S_{{n_1}B}^\gamma S_{{n_2}A}^x}}
       \mathrel{\left | {\vphantom {{S_{nA}^\gamma S_{{n_1}B}^\gamma S_{{n_2}A}^x} {S_{m\alpha }^x}}}
         \right. \kern-\nulldelimiterspace}
               {{S_{m\alpha }^x}} \right\rangle _{i{\omega _n}}}
     & \simeq & \alpha c_{AB}^\gamma \left( {{{\bf{R}}_{{n_1}}} - {{\bf{R}}_n}} \right)
     \nonumber \\ & & \times
     {\left\langle {{S_{{n_2}A}^x}}
       \mathrel{\left | {\vphantom {{S_{{n_2}A}^x} {S_{m\alpha }^x}}}
         \right. \kern-\nulldelimiterspace}
               {{S_{m\alpha }^x}} \right\rangle _{i{\omega _n}}},
     \label{eq:decoupling2}     
     \eea
       where $\alpha$ is a correction parameter	in this	approximation\cite{Kondo1972}
       to be determined from the constraint.
       The approximation is more accurate when one applies
       this decoupling scheme at Green's functions with more spins
       or in higher spatial dimensions\cite{Sasamoto2024}.
      Here, the correlation functions are defined by
      \be
      c_{{\alpha _1}{\alpha _2}}^\gamma \left( {{{\bf{R}}_{{n_2}}} - {{\bf{R}}_{{n_1}}}} \right) = \left\langle {S_{{n_1}{\alpha _1}}^\gamma S_{{n_2}{\alpha _2}}^\gamma } \right\rangle
      \ee
      with ${\bf R}_n$ being the coordinate vector of the $n$-th unit cell.

      After a tedious but straightforward calculation followed by a Fourier transform to momentum space, we obtain
      \be
        \left[ {{{\left( {i{\omega _n}} \right)}^2} - \frac{K^2}{2}} \right]
             {{G}^x}\left( {{\bf{q}},i{\omega _n}} \right)
             - {K^2}M\left( {\bf{q}} \right){{G}^x}\left( {{\bf{q}},i{\omega _n}} \right)
             \simeq KN \left( {\bf{q}} \right).
             \label{eq:G_eq}
             \ee
             The matrix form of the Green's function, ${{G}^x}\left( {{\bf{q}}, i{\omega _n}} \right)$,
             is given by:             
      \be
      {{G}^x}\left( {{\bf{q}},i{\omega _n}} \right) = \left( {\begin{array}{*{20}{c}}
{G_{AA}^x\left( {{\bf{q}},i{\omega _n}} \right)}&{G_{AB}^x\left( {{\bf{q}},i{\omega _n}} \right)}\\
{G_{BA}^x\left( {{\bf{q}},i{\omega _n}} \right)}&{G_{BB}^x\left( {{\bf{q}},i{\omega _n}} \right)}
      \end{array}} \right).
      \label{eq:G_matrix}
      \ee
      The components of the matrices $M({\bf q})$ and $N({\bf q})$ are:      
      \bea
          {M_{AA}}\left( {\bf{q}} \right)/\alpha
          &=& c_{AB}^y\left( 0 \right){e^{i{\bf{q}} \cdot {{\bf{n}}_1}}} + c_{AB}^z\left( {{{\bf{n}}_2}} \right){e^{i{\bf{q}} \cdot \left( {{{\bf{n}}_1} - {{\bf{n}}_2}} \right)}}, \\
          {M_{AB}}\left( {\bf{q}} \right)/\alpha
          &=&  - c_{AA}^y\left( { - {{\bf{n}}_2}} \right)
          - \left[ {c_{AB}^z\left( 0 \right)
              + c_{AB}^y\left( {{{\bf{n}}_2}} \right)} \right]{e^{ - i{\bf{q}} \cdot {{\bf{n}}_1}}}
          \nonumber \\
          & & - c_{AA}^z\left( {{{\bf{n}}_2}} \right){e^{ - i{\bf{q}} \cdot {{\bf{n}}_2}}}, \\
          {M_{BA}}\left( {\bf{q}} \right)/\alpha
          &=&  - c_{BB}^y\left( {{{\bf{n}}_2}} \right) - \left[ {c_{BA}^z\left( 0 \right) + c_{BA}^y\left( { - {{\bf{n}}_2}} \right)} \right]{e^{i{\bf{q}} \cdot {{\bf{n}}_1}}}
          \nonumber \\
          & & - c_{BB}^z\left( { - {{\bf{n}}_2}} \right){e^{i{\bf{q}} \cdot {{\bf{n}}_2}}}, \\
          {M_{BB}}\left( {\bf{q}} \right)/\alpha
          &=& c_{BA}^y\left( 0 \right){e^{ - i{\bf{q}} \cdot {{\bf{n}}_1}}}
          \nonumber \\ & & 
          + c_{BA}^z\left( { - {{\bf{n}}_2}} \right)
          {e^{ - i{\bf{q}} \cdot \left( {{{\bf{n}}_1} - {{\bf{n}}_2}} \right)}}.
          \eea
          and
          \bea
              {N_{AA}}\left( {\bf{q}} \right)
              &=& c_{AB}^z\left( 0 \right) + c_{AB}^y\left( {{{\bf{n}}_2}} \right)
              \nonumber \\
                        {N_{AB}}\left( {\bf{q}} \right)
                        &=&  - c_{AB}^y\left( 0 \right) - c_{AB}^z\left( {{{\bf{n}}_2}} \right){e^{ - i{{\bf{n}}_2} \cdot {\bf{q}}}}
                        \nonumber \\
                        {N_{BA}}\left( {\bf{q}} \right) &=&  - c_{BA}^y\left( 0 \right) - c_{BA}^z\left( { - {{\bf{n}}_2}} \right){e^{i{{\bf{n}}_2} \cdot {\bf{q}}}}
              \nonumber \\              
                        {N_{BB}}\left( {\bf{q}} \right)
                        &=& c_{BA}^z\left( 0 \right) + c_{BA}^y\left( { - {{\bf{n}}_2}} \right),
                        \eea
                        respectively.
                        We obtain a similar formula for ${{G}^y}\left( {{\bf{q}}, i{\omega _n}} \right)$
                        and ${{G}^z}\left( {{\bf{q}}, i{\omega _n}} \right)$.
                        However, due to the symmetry of the honeycomb lattice,
                        we do not need these for the isotropic case.

Solving Eq.~(\ref{eq:G_eq}), we obtain
\bea
    {G^x}\left( {{\bf{q}},i{\omega _n}} \right)
    &=& \frac{K}{{\left[ {{{\left( {i{\omega _n}} \right)}^2} - {{\left( {\omega _{\bf{q}}^{\left(  +  \right)}} \right)}^2}} \right]\left[ {{{\left( {i{\omega _n}} \right)}^2} - {{\left( {\omega _{\bf{q}}^{\left(  -  \right)}} \right)}^2}} \right]}}
    \nonumber \\
    & & \hspace{-7em} \times \left( {\begin{array}{*{20}{c}}
        {{{\left( {i{\omega _n}} \right)}^2} - {K^2}/2 - a_{\bf{q}}^*}&{{b_{\bf{q}}}}\\
        {b_{\bf{q}}^*}&{{{\left( {i{\omega _n}} \right)}^2} - {K^2}/2 - {a_{\bf{q}}}}
    \end{array}} \right){N_{\bf{q}}},
    \eea
    The energy dispersion $\omega _{\bf{q}}^{(\pm)}$ is given by    
    \be
    \omega _{\bf{q}}^{\left(  \pm  \right)} =
    \sqrt {\frac{{{K^2}}}{2} + {\mathop{\rm Re}\nolimits} {a_{\bf{q}}} \pm {\lambda _{\bf{q}}}},
    \ee
    with
    ${\lambda _{\bf{q}}} = \sqrt {{{\left| {{b_{\bf{q}}}} \right|}^2} - {{\left| {{\mathop{\rm Im}\nolimits} {a_{\bf{q}}}} \right|}^2}}$.
The terms $a_{\bf q}$ and $b_{\bf q}$ are defined by       
\be
   {a_{\bf{q}}} =  + \alpha {K^2}{e^{i{\bf{q}} \cdot {{\bf{n}}_1}}}\left[ {c_{AB}^y\left( 0 \right) + c_{AB}^z\left( {{{\bf{n}}_2}} \right){e^{ - i{\bf{q}} \cdot {{\bf{n}}_2}}}} \right],
   \ee
\bea
    {b_{\bf{q}}}
    &=&  - \alpha {K^2}\left[
      c_{AB}^z\left( 0 \right) {e^{ - i{\bf{q}} \cdot {{\bf{n}}_1}}}
      + c_{AB}^y\left( {{{\bf{n}}_2}} \right) {e^{ - i{\bf{q}} \cdot {{\bf{n}}_1}}}
      \right. \nonumber \\
      & & \left. 
        + c_{AA}^y\left( { - {{\bf{n}}_2}} \right)
        + c_{AA}^z\left( {{{\bf{n}}_2}} \right){e^{ - i{\bf{q}} \cdot {{\bf{n}}_2}}} \right],
    \eea
    
    In our Green's function approach, we determine the correlation functions in a self-consistent manner.
    The correlation functions are expressed in terms of the Green's function as
\be
c_{{\alpha _1}{\alpha _2}}^x \left( {\bf{r}} \right) =  - \frac{1}{N}\sum\limits_{\bf{q}} {{e^{i{\bf{q}} \cdot {\bf{r}}}}} {\left[ {{G^x }\left( {{\bf{q}},\tau  = {0^ + }} \right)} \right]_{{\alpha _1}{\alpha _2}}},
\ee
where $N$ is the number of unit cells.
In the numerical calculations below, we take $N = 45 \times 45$.
There are relationships between the correlation functions due to the absence of
magnetic long-range order and any symmetry breaking.
By utilizing these symmetries, we define
\be
   {c_1} \equiv c_{AB}^z\left( 0 \right) = c_{AB}^x\left( {{{\bf{n}}_1}} \right)
   = c_{AB}^y\left( {{{\bf{n}}_2}} \right),
\ee
\bea
    {c^{\prime}_1} & \equiv & c_{AB}^x\left( 0 \right) = c_{AB}^y\left( {{{\bf{n}}_1}} \right)
    = c_{AB}^z\left( {{{\bf{n}}_2}} \right) \nonumber \\
     &=& c_{AB}^y\left( 0 \right) = c_{AB}^z\left( {{{\bf{n}}_1}} \right) = c_{AB}^x\left( {{{\bf{n}}_2}} \right),
\eea
\bea
    {c_2} & \equiv & c_{AA}^y\left( {{{\bf{n}}_2}} \right) = c_{AA}^x\left( {{{\bf{n}}_1} - {{\bf{n}}_2}} \right) = c_{AA}^z\left( {{{\bf{n}}_1}} \right) \nonumber \\
    &=& c_{AA}^z\left( {{{\bf{n}}_2}} \right) = c_{AA}^y\left( {{{\bf{n}}_1} - {{\bf{n}}_2}} \right) = c_{AA}^x\left( {{{\bf{n}}_1}} \right).
    \eea
    The self-consistent equations to be solved are then given by
\bea
    {c_1} &=& \frac{1}{{\beta N}}\sum\limits_{\bf{q}} {\sum\limits_{s =  \pm }
      {{e^{i{\bf{q}} \cdot {{\bf{n}}_1}}}} }
    \frac{{Z_{AB}^s}}{{{{\left( {\omega _{\bf{q}}^s} \right)}^2}}}
    g\left( {\frac{{\beta \omega _{\bf{q}}^s}}{2}} \right)
    \\
    c_1^{\prime}
    &=& \frac{1}{{\beta N}}\sum\limits_{\bf{q}} {\sum\limits_{s =  \pm }
        {\frac{{Z_{AB}^s}}{{{{\left( {\omega _{\bf{q}}^s} \right)}^2}}}
          g\left( {\frac{{\beta \omega _{\bf{q}}^s}}{2}} \right)} }
      \\
        {c_2} &=& \frac{1}{{\beta N}}\sum\limits_{\bf{q}} {\sum\limits_{s =  \pm }
          {{e^{i{\bf{q}} \cdot \left( {{{\bf{n}}_1} - {{\bf{n}}_2}} \right)}}} }
        \frac{{Z_{AA}^s}}{{{{\left( {\omega _{\bf{q}}^s} \right)}^2}}}
        g\left( {\frac{{\beta \omega _{\bf{q}}^s}}{2}} \right)
        \\
        \frac{1}{4} &=& \frac{1}{{\beta N}}\sum\limits_{\bf{q}} {\sum\limits_{s =  \pm }
          {\frac{{Z_{AA}^s}}{{{{\left( {\omega _{\bf{q}}^s} \right)}^2}}}
            g\left( {\frac{{\beta \omega _{\bf{q}}^s}}{2}} \right)} }
        \eea
        where $g(x) = x \coth x$.
        The last equation is obtained from
        $c_{AA}^x \left( 0 \right) = 1/4$.
        $Z_{AA}^ \pm$ and $Z_{AB}^ \pm$ are given by
        \be
        Z_{AA}^ \pm  =  \pm
        \frac{2\left( {i {\rm Im} {a_{\bf{q}}}
              \pm {\lambda _{\bf{q}}}} \right){c_1} + {b_{\bf{q}}}\left( {{e^{i{{\bf{n}}_2}
                  \cdot {\bf{q}}}} + 1} \right)
            {c_1^{\prime}}}{{2{\lambda _{\bf{q}}}}},
        \ee
        \be
        Z_{AB}^ \pm  =  \mp \frac{{\left( {{e^{ - i{{\bf{n}}_2} \cdot {\bf{q}}}} + 1} \right)
            \left( {i{\rm Im} {a_{\bf{q}}} \pm {\lambda _{\bf{q}}}} \right){c_1^{\prime}} + 2{b_{\bf{q}}}{c_1}}}{{2{\lambda _{\bf{q}}}}}.
        \ee

\begin{figure}[htbp]
  \includegraphics[width=0.9 \linewidth, angle=0]{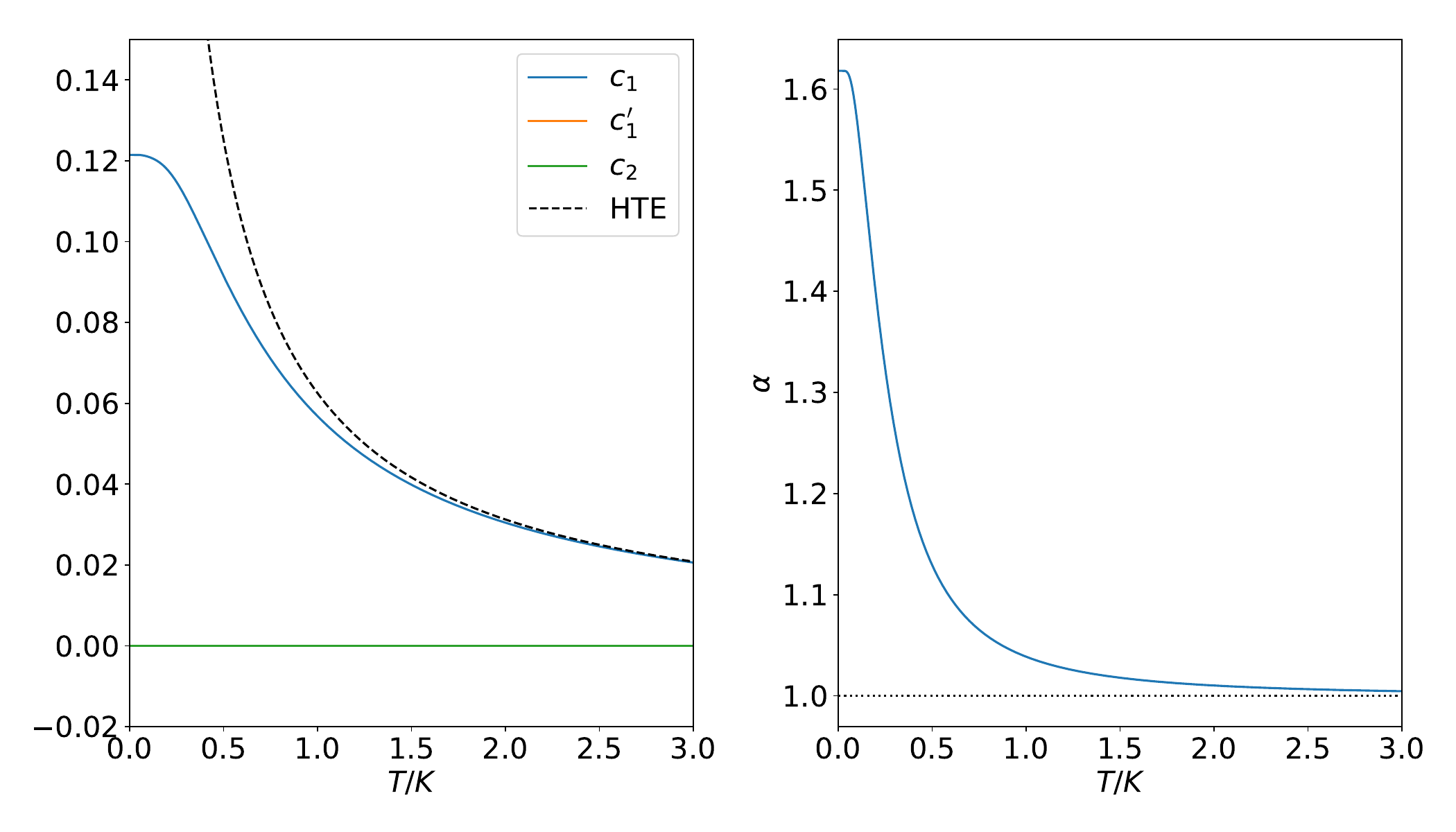}
  \caption{
    \label{fig:c1_alpha}
    (Color online)
    Temperature dependence of the correlation functions and the correction parameter $\alpha$. 
    Left panel: The temperature dependencies of $c_1$, $c^{\prime}_1$, and $c_2$ are shown.
    The value of $4c_1$ approaches the exact theoretical value\cite{Baskaran2007} of 0.5249 at $T=0$
    and
      aligns with high-temperature expansion result, which is indicated by the dashed line,
      at elevated temperatures,
    while $c^{\prime}_1$ and $c_2$ remain zero across the temperature range.
    Right panel: The variation of the parameter $\alpha$ with temperature,
    demonstrating the exactness of the decoupling approximation at high temperatures
    (where $\alpha$ approaches 1) and the necessity of this correction parameter
    at lower temperatures, where $\alpha$ deviates from 1.
  }
\end{figure}

The temperature dependencies of $c_1$, $c^{\prime}_1$, and $c_2$ are shown
in the left panel of Fig.~\ref{fig:c1_alpha}.
The value of $4c_1$ converges to 0.4859
as $T$ approaches zero, closely approximating the exact value\cite{Baskaran2007} of
$0.5249 \equiv 4c_1^{\rm exact}$.
At high temperatures, the behavior of $c_1$ aligns well with the high-temperature expansion result,
expressed as $c_1 = \beta K/16 + O(\beta^3)$.
Meanwhile, $c^{\prime}_1$ and $c_2$ remain zero
across all temperatures, a finding rigorously confirmed\cite{Baskaran2007} at $T=0$ 
and consistent even at finite temperatures.
The vanishing of $c^{\prime}_1$ and $c_2$ can be understood as follows.
Consider, for instance, $c_{AB}^x(0) = \left\langle S_{0A}^x S_{0B}^x \right\rangle$,
which corresponds to $c^{\prime}_1$.
In this correlation function, the spin operator $S_{0B}^x$ flips the $\mathbb{Z}_2$ flux values
of the adjacent hexagons that share the $x$-bond emanating from the B sublattice in the 0-th unit cell.
Similarly, the spin operator $S_{0A}^x$ flips the $\mathbb{Z}_2$ flux values of the adjacent hexagons
that share the $x$-bond emanating from the A sublattice in the 0-th unit cell.
Since the hexagons affected by $S_{0B}^x$ and $S_{0A}^x$ are different
and the Hamiltonian $H$ does not alter the $\mathbb{Z}_2$ flux values,
the thermal average of their product is identically zero.
This reasoning also applies to the other correlation functions,
highlighting a unique characteristic of the Kitaev model.
  The same argument leads to
  $\left\langle {
    {T_\tau }
    S_{n {\alpha_1} }^{\gamma_1} \left( \tau  \right)
    S_{m {\alpha_2} }^{\gamma_2} \left( 0 \right)
  } \right\rangle \equiv 0$
  for $\gamma_1 \neq \gamma_2$.
  The flux values flipped by $S_{m{\alpha _2}}^{\gamma_2}$ at imaginary time 0
  cannot be flipped again by $S_{n{\alpha _1}}^{\gamma_1}$ at imaginary time $\tau$.
  We note that this result is exact.
  From this observation, we may restrict the two-spin Green's function
  to the form of Eq.~(\ref{eq:G_spin:real_space}).
We also note that the vanishing of $c^{\prime}_1$ and $c_2$
results in $\omega _{\bf{q}}^{\pm}$ forming a flat band,
given by $K\sqrt{\left(1 \pm 4\alpha {c_1}\right)/2}$.

In the right panel of Fig.~\ref{fig:c1_alpha},
we present the temperature dependence of $\alpha$.
As temperature increases, $\alpha$ approaches one,
indicating that the decoupling approximation in Eqs.~(\ref{eq:decoupling1}) and (\ref{eq:decoupling2})
becomes exact without the need for the correction parameter $\alpha$.
However, deviations of $\alpha$ from one at lower temperatures suggest
the necessity of this correction parameter in the equations.
%-------------------------------------------------------------------
%-------------------------------------------------------------------

\begin{figure}[htbp]
  \includegraphics[width=0.9 \linewidth, angle=0]{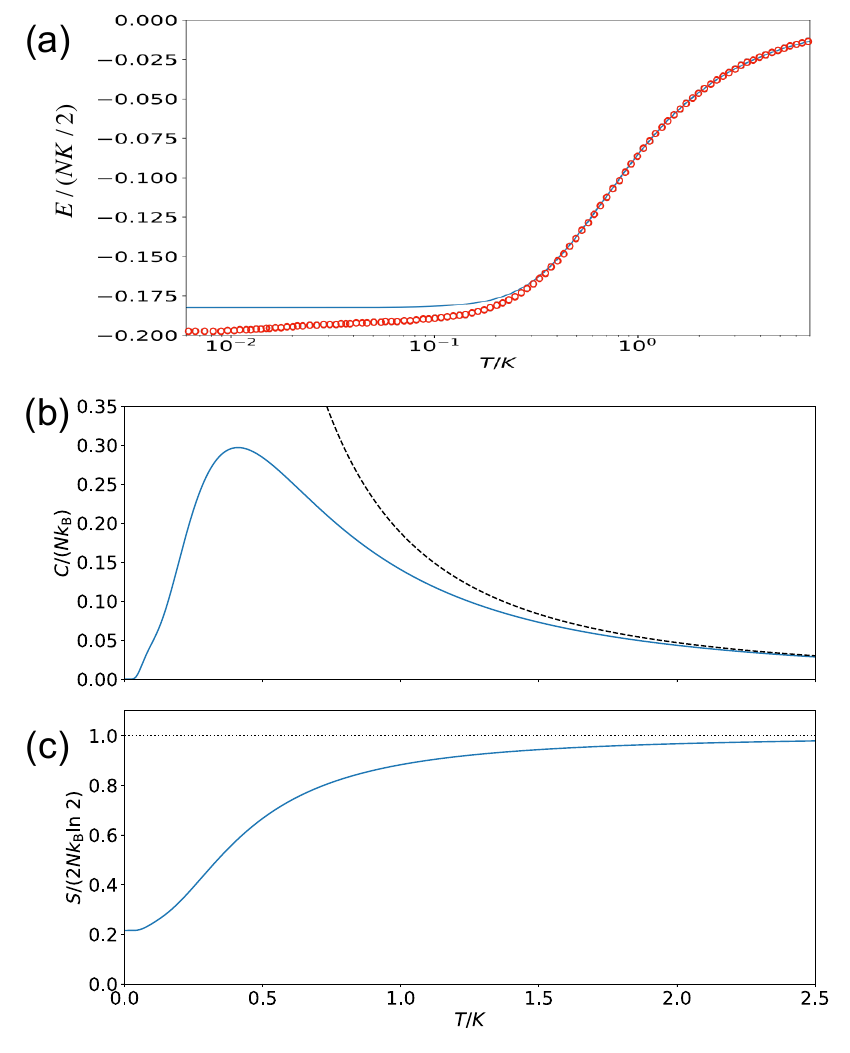}
  \caption{
    \label{fig:E_C_S}
    (Color online)
    Temperature dependencies of (a) the energy $E = -3 N K c_1/4$, (b) specific heat, and (c) entropy.
    The red circles in (a) are data taken from Ref.~\onlinecite{Motome2020}.
    The specific heat, calculated from the temperature derivative of the internal energy $E$,
    is compared with the high-temperature expansion result $C_{\rm HTE} = 3N K^2 \beta^2/16$.
    It displays a broad peak around $T/K \simeq 0.4$ without any lower-temperature peaks.
    The entropy is computed based on Eq.~(\ref{eq:entropy}).
    }
\end{figure}

From the temperature dependence of $c_1$, the internal energy can be computed using the relation
\be
E = \left\langle H \right\rangle = - \frac{3}{4}N K {c_1}.
\ee
We compared the temperature dependence of the energy with the results shown in Fig.~15(a)
of Ref.~\onlinecite{Motome2020}, as shown in Fig.~\ref{fig:E_C_S}(a).
We found an almost perfect match for $T > T_H$ with $T_H \simeq 0.375K$,
but our energy values are slightly higher for $T < T_H$.
This discrepancy likely originates from the error in the $c_1$ value at $T=0$.
Although the relative error is
$|(c_1^{\rm exact}-c_1)/c_1^{\rm exact}| \simeq 0.07$,
minor details and significant physics, such as the fractionalization of spins,
may be hidden within this discrepancy.

Figure~\ref{fig:E_C_S} shows the temperature dependencies of the specific heat and entropy. 
The specific heat is calculated from the temperature derivative of the internal energy $E$ 
and compared with the high-temperature expansion (HTE) result,
$C_{\rm HTE} = 3N K^2 \beta^2/16$. 
The entropy is computed using the formula:
\be
S = S\left( \infty \right) - \int_T^{T_0} dT \frac{C(T)}{T} - \int_{T_0}^\infty dT \frac{C_{\rm HTE}(T)}{T},
\label{eq:entropy}
\ee
where $S\left( \infty \right) = 2N\ln 2$ and $T_0$ is a sufficiently large value 
at which the Green's function result for the specific heat
closely approximates the HTE result (here, we take $T_0/K = 10$). 
The specific heat exhibits a single broad peak around $T/K \simeq 0.4$, 
consistent with mean field calculations \cite{Saheli2024}. 
We likely fail to observe the low-temperature peak around $T_L \simeq 0.012 K$ reported
in Ref.~\onlinecite{Nasu2015} due to the discrepancy mentioned above. 
The entropy indicates the presence of residual entropy, 
but if fractionalization were captured, this residual entropy should vanish, 
as it has been rigorously shown that there are no $\mathbb{Z}_2$ fluxes 
in the ground state and the system must satisfy the third law of thermodynamics.
%-------------------------------------------------------------------
%-------------------------------------------------------------------

  In conclusion, we have investigated the Kitaev model at finite temperatures
  using a Green's function approach that maintains SU(2) symmetry.
  Our computations of the temperature dependence of the correlation functions revealed
  that the nearest neighbor spin-spin correlation functions closely match the exact values
  at zero temperature.
  Additionally, we have presented several exact results concerning the spin Green's function
  and spin-spin correlation functions specific to the Kitaev model.
  These findings enhance our understanding of the Kitaev model and provide a novel approach
  to this model without relying on Majorana fermions.
  This work opens up new avenues for exploring the dynamic properties of the Kitaev model
  and its applications in the study of quantum spin liquids and topological quantum computing.

%------------------------------------------------------------------------
\begin{acknowledgments}
  The authors thank K. Harada and D. Sasamoto for helpful discussions.
\end{acknowledgments}

\bibliography{../../../../refs/lib}
\end{document}